\documentclass[%
 reprint,
 amsmath,amssymb,
 aps,
pra]{revtex4-1}

\usepackage[utf8]{inputenc}
\usepackage{graphicx}
\usepackage{dcolumn}
\usepackage{bm}
\usepackage{ulem}
\usepackage{xcolor}
\usepackage{lipsum}
\usepackage{hyperref}
\hypersetup{
    colorlinks=true,
    linkcolor=blue,
    filecolor=magenta,      
    urlcolor=black,
    citecolor=blue,
    pdftitle={Overleaf Example},
    pdfpagemode=FullScreen,
    }

\begin{document}

\preprint{APS/123-QED}

\title{Accurate and precise optical phase sensor based on a non-linear quantum Sagnac interferometer}

\author{Romain Dalidet, Laurent Labonté, Gregory Sauder, Sébastien Tanzilli, Anthony Martin}
\email{anthony.martin@univ-cotedazur.fr}
\affiliation{Université Côte d’Azur, CNRS, Institut de physique de Nice, France}

\begin{abstract}
Optical phase measurements play a key role in the detection of macroscopic parameters such as position, velocity, and displacement. They also permit to qualify the microscopic properties of photonic waveguides such as polarization mode dispersion, refractive index difference, and chromatic dispersion. In the quest for ever-better measurement performance and relevance, we report an original quantum non-linear interferometer based on a Sagnac configuration allowing precise, accurate, self-stabilized, and reproductible optical phase measurement. The potential of this system is demonstrated through the measurement of second-order dispersion, namely chromatic dispersion, of a commercial dispersion-shifted fiber at telecommunication wavelength. We assess precision by exhibiting a statistical error of $7.10^{-3}\, \%$, showing more that one order of magnitude compares to state-of-the-art measurements. Additionally, the accuracy of the second-order dispersion value is determined through the measurement of the third-order dispersion, showing a quadratic error as low as 5\,\%. Our system promises the development of photonic-based sensors enabling the measurements of optical-material properties in a user-friendly manner.
 \end{abstract}

\maketitle

\section{Introduction}
Precise measurement is at the heart of science and technology~\cite{daryanoosh_experimental_2018}. A fundamental concern lies in achieving the best precision in measuring optical phase, which impacts a large variety of domains encompassing gravitational wave detection,microscopy, and optical coherence tomography~\cite{jia_ligo_2024, ndagano_quantum_2022, abouraddy_quantum-optical_2002}. Quantum physics allows, in principle, ultimate precision achievable by exploiting correlated quantum resources, such as entangled states, offering significant advantages over even the most optimized iterations of classical counterparts~\cite{giovannetti_quantum-enhanced_2004, moreau_demonstrating_2017}. Quantum-enhanced optical phase estimation holds the promise of substantial improvements in all interferometry-based measurements~\cite{genovese_experimental_2021}. The central challenge, however, remains the stability of the interferometer itself. Environmental factors, such as temperature fluctuations and vibrations, introduce phase drifts that must be smaller than the variations being measured. Most of the interferometric phase measurements are based on Mach-Zehnder or Michelson interferometer architectures which require the use of sophisticated stabilization systems. The associated experimental complexity merely shifts the underlying challenge without fully resolving it~\cite{Grassani:14}.\\
In contrast, Sagnac interferometers are immune to phase drifts, as reciprocal phase shifts cancel out when the drift-rate is slower than the light propagation time, resulting in zero relative phase. This characteristic has greatly benefited the measurement of non-reciprocal phenomena, such as Faraday rotation and the gyroscopic effect, in both classical~\cite{Culshaw_2006} and quantum~\cite{fink_entanglement-enhanced_2019} regimes. Conversely, this configuration also finds limitations as Sagnac interferometers are inherently insensitive to chromatic dispersion (CD), a critical parameter for numerous photonics applications where both high precision and stability are essential~\cite{agrawal_nonlinear_1995, fasel_quantum_2004}.\\
In this paper, we introduce an original interferometric architecture that integrates a novel type of non-reciprocity by embedding a non-linear element within the Sagnac loop. The latter grants access to previously unattainable parameters, such as the optical phase in the frequency domain, enabling precise CD measurements of the sample under test. We demonstrate several significant advantages: (i) the inherent system self-stabilization facilitated by a common path for all beams, (ii) a deterministic output for photons exiting the interferometer, unlike the typical 50\% output~\cite{shi_generation_2004}, thereby reducing the overall loss budget of the system, and (iii) a straightforward alignment achieved through a fully fibered configuration operating at telecom wavelengths. Furthermore, the use of energy–time entangled photon pairs and coincidence counting to measure spectral correlation functions leads to non-local dispersion cancellation and the suppression of systematic errors~\cite{kaiser_quantum_2018}. These advances collectively push the boundaries of precision and accuracy in CD measurements, positioning our approach at the forefront of both classical and quantum measurement techniques~\cite{fasel_quantum_2004, kaiser_quantum_2018, dalidet24}.

\section{Theory}

\begin{figure}[h]
    \centering
    \includegraphics[width=1\linewidth]{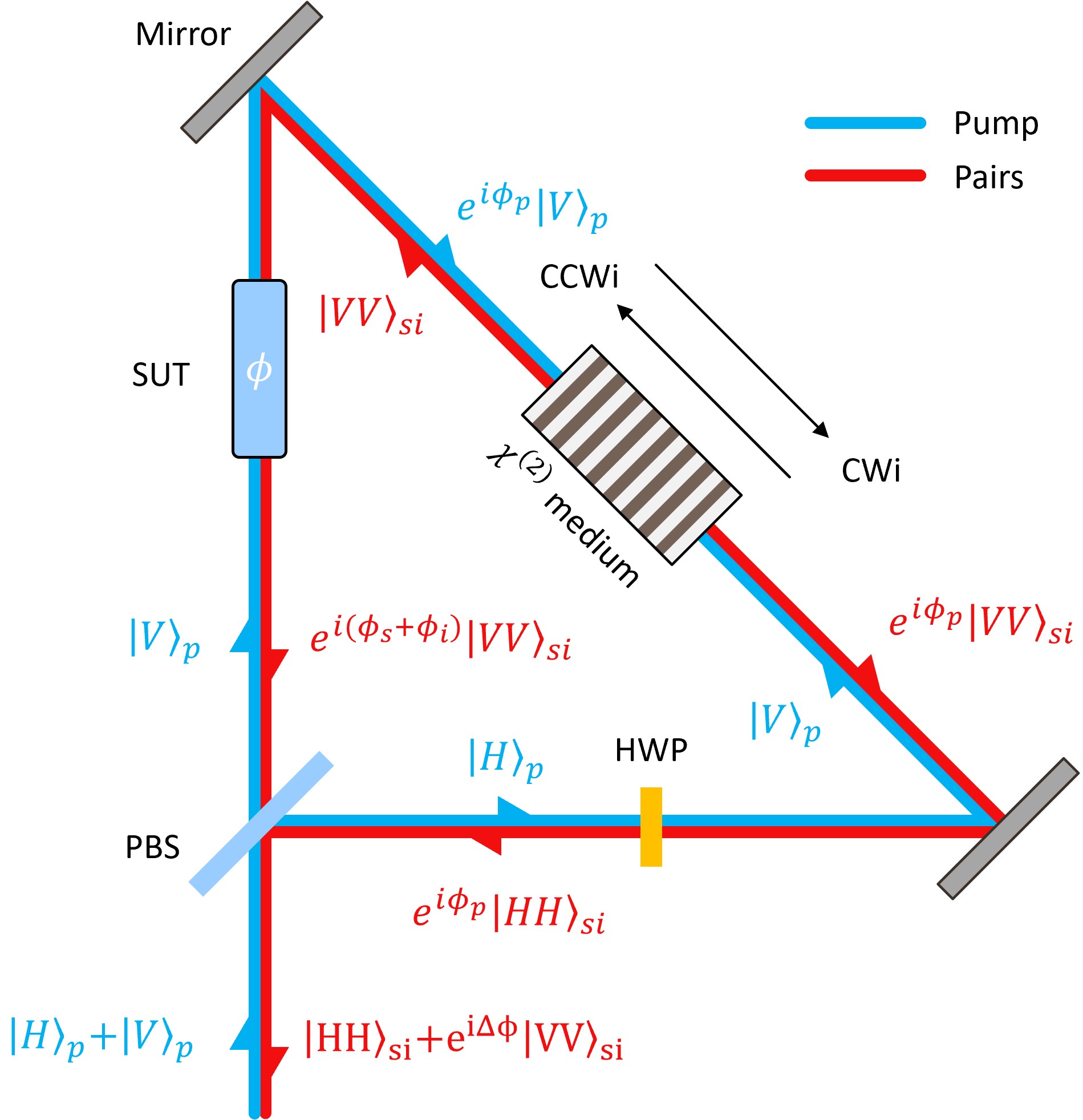}
    \caption{Principle of a non-linear Sagnac interferometer in free space configuration using a type-0 phase matched medium. The blue (red) path represents the propagation of the pump (pairs). HWP: half-wave plate; PBS: polarizing beam splitter; CWi: clockwise; CCWi: counter-clockwise; SUT: sample under test.}
    \label{fig: bulk configuration}
\end{figure}

The schematic of the quantum non-linear Sagnac interferometer is shown in Fig.\ref{fig: bulk configuration}. It comprises a common-path bulk configuration using a polarization beam splitter (PBS). The loop itself consists of a half-wave plate (HWP), a second-order non-linear medium, and a dispersive medium acting as sample under test (SUT). We assume that the path-difference between the non-linear medium and the PBS in both clockwise (CWi) and counterclockwise (CCWi) only comes from the dispersive medium. 
For the sake of clarity, we first consider the pump at frequency $\omega_p$ to be vertically polarized. The latter propagates in the CWi path through the SUT and accumulates a phase denoted $\phi_p$. Subsequently, within the non-linear medium, it undergoes spontaneous parametric down conversion (SPDC) to generate photon pairs at the degenerate frequency $\omega_0 = \frac{\omega_p}{2}$. A type-0 phase matching condition is achieved, ensuring polarization conservation between the pump and the generated paired photons. The latter then pass through the half-wave plate (HWP) to rotate their polarization to horizontal state. Finally, they exit  through the pump arm, in opposite direction. As SPDC preserves the phase, the final state in the CWi path reads:

\begin{equation}\label{eq: CW state}
    |\psi_{cwi}\rangle_{si} = e^{i\phi_p}|HH\rangle_{si}\, ,
\end{equation}
where the subscripts $s$ and $i$ denote down-converted signal and idler photons. Conversely, if the pump is horizontally polarized, it propagates through the CCWi path. 
The pump light first travels through the HWP to rotate its polarization and then undergoes down-conversion into the crystal. The paired photons pass through the SUT and exit the PBS through the pump arm. Thus, the final state in the CCWi path is given by:

\begin{equation}\label{eq: CCW state}
    |\psi_{ccwi}\rangle_{si} = e^{i\phi_s+\phi_i}|VV\rangle_{si}\, .
\end{equation}
Finally, we consider the pump to be diagonally polarized:

\begin{equation}\label{eq: pump initial state}
    |\psi_{in}\rangle_p = |D\rangle_p \equiv \frac{1}{\sqrt{2}}\bigl(|H\rangle_p + |V\rangle_p\bigr) \, ,
\end{equation}
where the subscript $p$ denotes the pump. No relative phase between the horizontal and vertical polarization components is considered. Assuming that the SPDC efficiency is equal in both directions, the final state after propagation is the coherent superposition of equation \ref{eq: CW state} and \ref{eq: CCW state} :

\begin{align}
    |\psi_{out}\rangle_{si} & =  \frac{1}{\sqrt{2}}\Bigl[|HH\rangle_{si} + e^{i\Delta\phi}|VV\rangle_{si}\Bigr]\label{eq: max polar entangled} \, ,  \\
    \Delta\phi & = \phi_s + \phi_i - \phi_p = (k_s+k_i-k_p)L\, .
\end{align}
Here $k$ and $L$ denote the wavevector of the interacting photons and the length of the dispersive medium, respectively. Therefore, the photons are prepared in a maximally polarized entangled state, and the relative phase between the contributions to the state solely depends on the dispersive properties of the SUT. The phase imparted by the dispersive medium can be expressed as a Taylor expansion of the wavevector of the signal and idler photons around the degenerate frequency:

\begin{equation}
    \frac{\Delta\phi}{L}= -k_p + \sum_{k=i,s}\sum_{n=0}^{\infty}\frac{\Delta\omega^n_k}{n!} \beta^{(n)} \, ,
    \label{eq_phi_taylor}
\end{equation}
where $\beta^{(n)}=\left.\frac{\partial^n k}{\partial \omega^n}\right|_{\omega_0}$ and $\Delta\omega$ stands for the detuning from $\omega_0$. The energy conservation of the non-linear process implies $\Delta\omega_s = -\Delta\omega_i$. Consequently, every odd-order term of the dispersion vanishes, a phenomenon called non-local dispersion cancellation \cite{PhysRevA.45.3126_franson_cd_cancel}. By limiting the Taylor expansion to the second order, \textit{i.e.} neglecting fourth-order dispersion, the relative phase carried by the entangled state reads:

\begin{equation}\label{eq: phase difference}
    \begin{split}
        \Delta\phi & = (\beta^{(2)}\Delta\omega^2+2\beta^{(0)}-k_p)L \\
        & = \beta^{(2)}\Delta\omega^2L + \phi_{off} \,.
    \end{split}
\end{equation}

\begin{figure*}[!ht]
    \centering
    \includegraphics[width=0.85\linewidth]{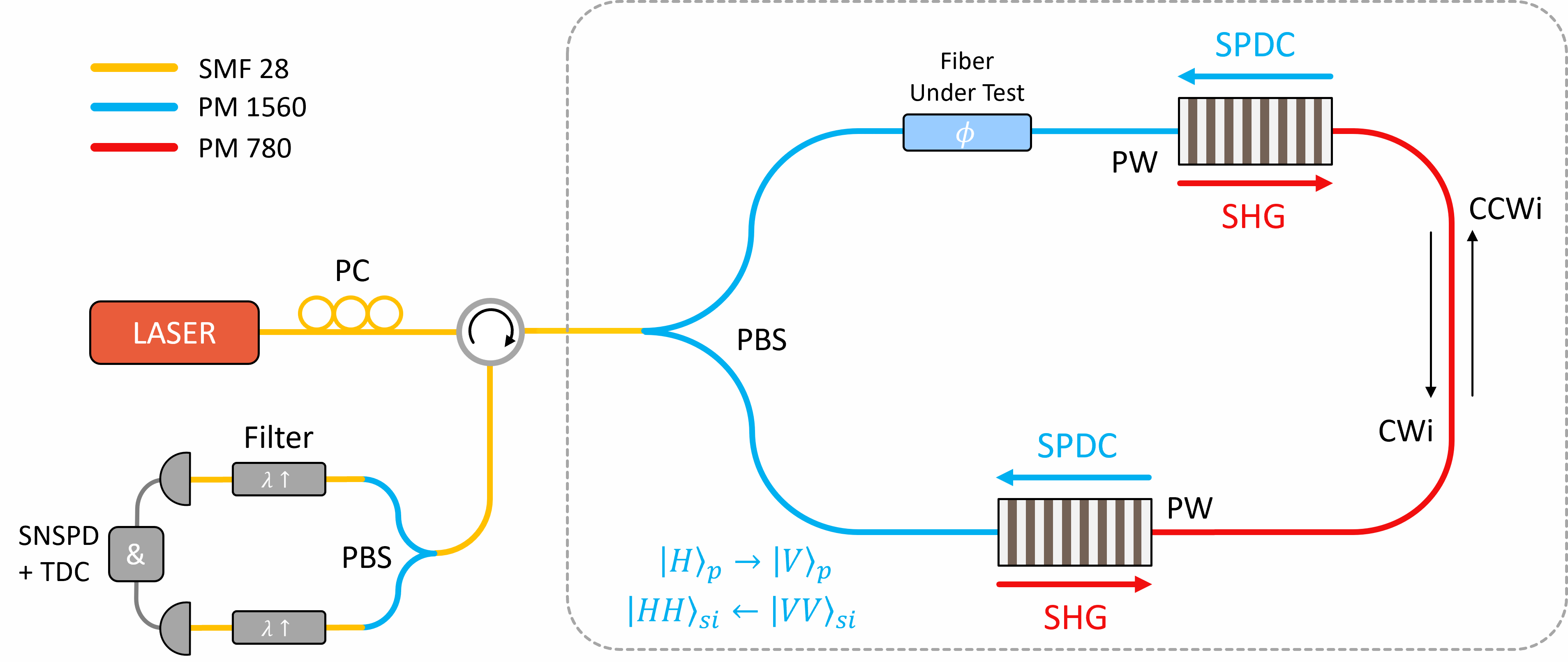}
    \caption{Experimental setup for CD measurement using the non-linear Sagnac interferometer. An amplified and filtered telecom laser is injected into the Sagnac loop, which consists of the SUT and of two type-0 periodically poled lithium niobate (PPLN) waveguides. The blue (red) path represents polarization maintaining telecom (visible) fiber. The two ouput arms of the PBS are aligned to the same polarization axis such that only one polarization state propagates inside the interferometer. At the loop's output, a circulator directs the photons to the detection apparatus, comprising a PBS, two tunable bandpass filters, and single-photon detectors for recording coincidence measurements. PW: pigtailed waveguide. SNSPD: superconducting nanowire single-photon detector; TDC: time to digital converter.}
    \label{fig: Sagnac source 2}
\end{figure*}
where $\beta^{(2)}$ represents the CD. Therefore, two-photon interference pattern measured from the entangled state in the spectral domain will lead to quadratic fringes allowing the direct extraction of the CD. This striking feature of quantum white light interferometry (QWLI) stands as a unique and precise method for CD measurement. Here, it is the only free parameter of the relative phase, in opposition to classical white light interferometry where odd-order dispersion terms are still present \cite{Grosz:14_first_order}. Furthermore, the polarized Sagnac configuration offers two main advantages. On one hand, as the system is self-stabilized, it is extremely immune to environmental drifts (temperature, vibration) and does not require any passive or active stabilisation. It therefore enables the development of an easy-to-use, compact, and robust experimental setup. On the other hand, it relies on polarization entanglement, avoiding the inherent 3 dB loss attributed to time-energy entanglement in the case of a non-deterministic splitting~\cite{PhysRevLett.62.2205_3dblosses, oser_high-quality_2020}. While classical CD measurement, such as phase modulation \cite{Baker:14_CD_kerr} and time of flight \cite{PAGE2006161_TOD}, need dozens of meter to kilometric sample lengths, QWLI based on non-linear Sagnac interferometer provides the capability to measure CD in samples as short as a few centimeters. This distinctive feature opens up avenues for precisely characterizing the dispersion of various materials, including optical chips (such as microrings and silicon waveguides), bulk glasses, or non-linear media~\cite{bogaerts_silicon_2012}.\\

\section{Experimental Realization}

\textit{From theory to experiment.} To ensure compliance with telecom standards, an experimental setup based on cascaded up- and down-conversion is used \cite{Cabrejo-Ponce_2022}, as depicted in Fig.~\ref{fig: Sagnac source 2}, where the interferometric setup lies in the grey dotted line. In addition of being fully fibered, this configuration does not involve the use of expensive and cumbersome dual-wavelength components mandatory in free space experiments. It comprises a PBS, a fiber under test (SUT) and two cascaded type-0 pigtailed non-linear waveguides. 
Here, the initial pump state is that of Eq.~\ref{eq: pump initial state}, but the pump wavelength is adjusted to match the degenerate wavelength of the paired photons ($\omega_p = \omega_0$). Upon passing through the PBS, the pump beam is once again split into the CWi and CCWi paths, which corresponds to the decomposition of the diagonal state $|D\rangle$ in the $\{ |H\rangle, |V\rangle \}$ basis. Note that the two outputs of the PBS are precisely aligned to the same axis, ensuring that only vertically polarized light propagates within the loop. Within the CWi path, the pump first propagates trough the SUT, undergoes up-conversion in the first waveguide through second harmonic generation (SHG), and then is down-converted into photon pairs in the second waveguide. At the output, the quantum state associated with the CWi path reads:

\begin{equation}\label{eq: CW state 2}
    |\psi_{cwi}\rangle_{si} = e^{i2\phi_0}|HH\rangle_{si}\, ,
\end{equation}
where the factor 2 comes from the SHG process. Conversely, in CCWi path, the pump first undergoes up-conversion and then down-conversion, with the photon pairs passing through the SUT. In this case, the contribution to the quantum state of the CCWi path is given by Eq. \ref{eq: CCW state}. Finally, assuming that the non-linear processes are identical in both directions, the output state carried by the photons pairs is a maximally, polarization entangled state, as described in Eq.~\ref{eq: max polar entangled}. Here, the relative phase difference contains the dispersion cancellation property:

\begin{equation}\label{eq: phase difference 2}
     \Delta\phi = (\beta^{(2)}\Delta\omega^2+2\beta^{(0)}-2k_0)L \, .
\end{equation}
Note that the only difference from Eq.~\ref{eq: phase difference} is the value of the phase offset. 

\begin{figure}[!ht]
    \centering
    \includegraphics[width=1\linewidth]{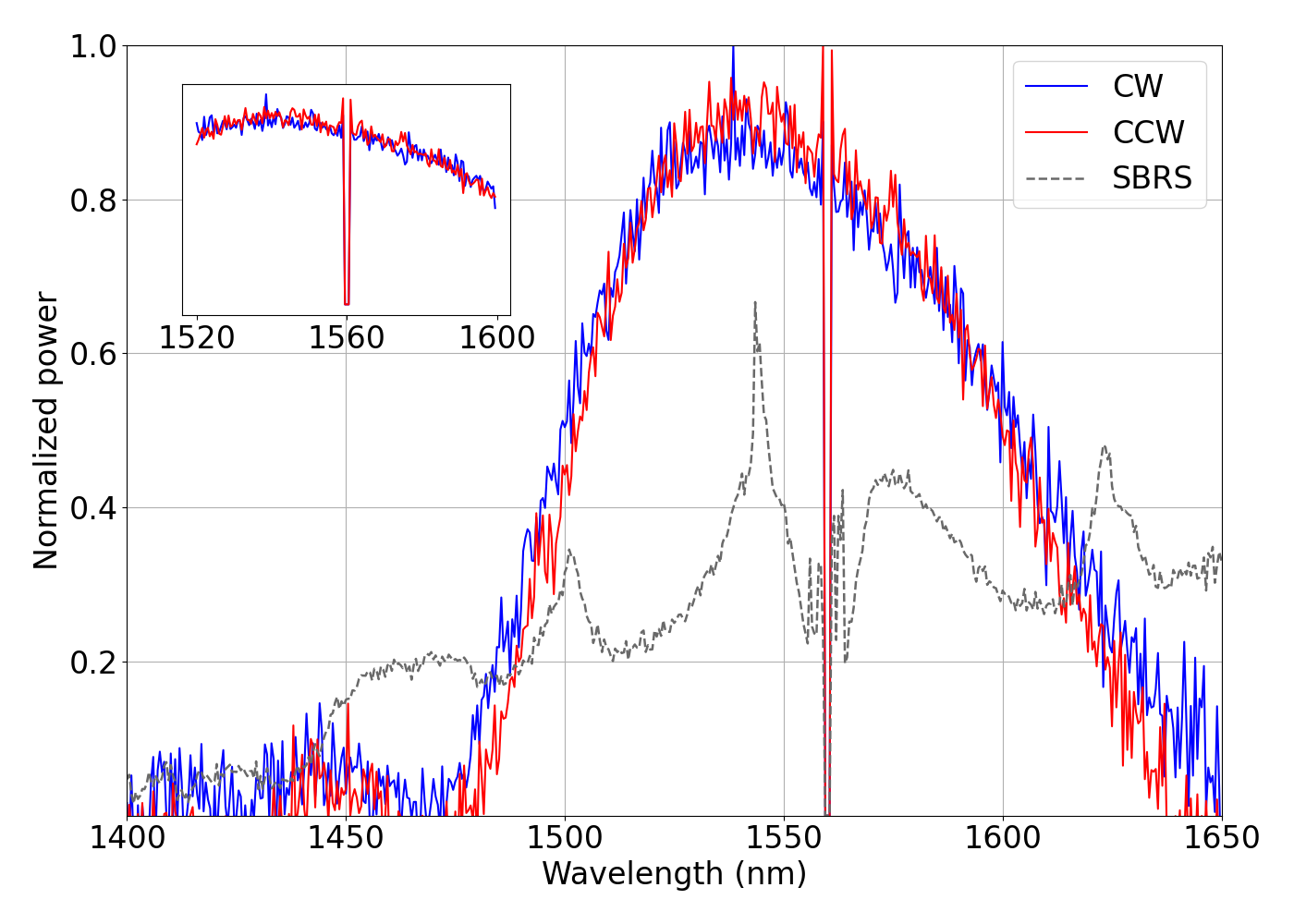}
    \caption{Spectrum of the photon pairs measured by inserting an optical spectrum analyser inside the Sagnac loop. The blue curve represents the spectrum in the CWi path, while the red curve represents the spectrum in the CCWi path. The inset provides a closer view of the measurement area. The grey dotted line corresponds to the Spontaneous Raman Back Scattering (SBRS) spectrum. The dip around 1560 nm corresponds to the wavelength-division multiplexers (WDMs) used to filter out the residual pump.}
    \label{fig: spectrum and raman}
\end{figure}

\textit{Description of the experimental apparatus.} A tunable telecom laser is first amplified and propagates in a sequence of filters, commonly referred to as wavelength dense multiplexers (WDMs), to eliminate amplified spontaneous emission. The polarization controller (PC) associated with the PBS allows the precise control of the infrared pump power to pre-compensate any discrepancies in brightness between the two directions. The laser light then enters the Sagnac loop, which consists of two nonlinear waveguides in periodically poled lithium niobate (PPLN). The use of polarization-maintaining fibers within the loop eliminates the need for active polarization control and optimizes the second-harmonic generation (SHG) and spontaneous parametric down-conversion (SPDC) processes. To ensure high visibility interference patterns, it is imperative that both the SPDC spectra and coincidence counts from each path are identical to prevent distinguishability. Fig.~\ref{fig: spectrum and raman} displays the measured spectra from the CWi and CCWi paths. By fine-tuning the temperature of both PPLN, a nearly perfect spectral overlap is achieved.  The SUT is a commercially available polarization maintaining dispersion-shifted fiber of length $L=0.9\, \text{m}$. At the output of the circulator, additional WDMs are used to prevent any residual pump light from entering the detection stage. This stage comprises a PC and a PBS to project the entangled state into a specific polarization basis. Each arm of the PBS is followed by a tunable bandpass filter, allowing for coincidence measurements in the spectral domain.

Finally, the measurement are performed using superconducting nanowire single-photon detectors and a time to digital converter to record coincidence events. Given the assumption of identical spectra for the photon pairs in both CWi and CCWi paths, the coincidence rate is determined by:

\begin{equation}\label{eq: coinc rate}
    P_c(\Delta\omega) \propto \frac{1}{2}sinc\bigl(\frac{\Delta k_c L_c}{2}\bigr)\bigl[1+Vcos(\Delta\phi) \bigr]\, ,
\end{equation}
where $\Delta k_c$ and $L_c$ are the phase matching  and length of the waveguides, $V$ stands as the visibility of the two photon interferences and $\Delta\phi$ is given by Eq.~\ref{eq: phase difference 2}. 
It is worth mentioning that spontaneous backward Raman scattering (SBRS) occurs within the Sagnac loop, as shown by the spectrum in Fig.~\ref{fig: spectrum and raman}. The dotted grey line was obtained with the crystals at room temperature to suppress SHG and SPDC processes during the SBRS measurement. The initial set of peaks around the central frequency is a result of SBRS within the fiber, while the subsequent set, located away from the central frequency, corresponds to SBRS within the crystals \cite{REPELIN1999819}. This phenomenon exhibits a linear dependence on both input power and material length \cite{Eraerds_2010}. However, photons generated via SBRS induce uncorrelated events and do not contribute directly to coincidence detection. A large number of SBRS photons can lead to accidental coincidences, thereby reducing the coincidence-to-accidental ratio. In our specific case, the proportion of accidental SBRS coincidences is less than 1\% of the measured SPDC coincidence peak, and is therefore negligible.

\section{Results}

\begin{figure}[!ht]
    \centering
    \includegraphics[width=0.9\linewidth]{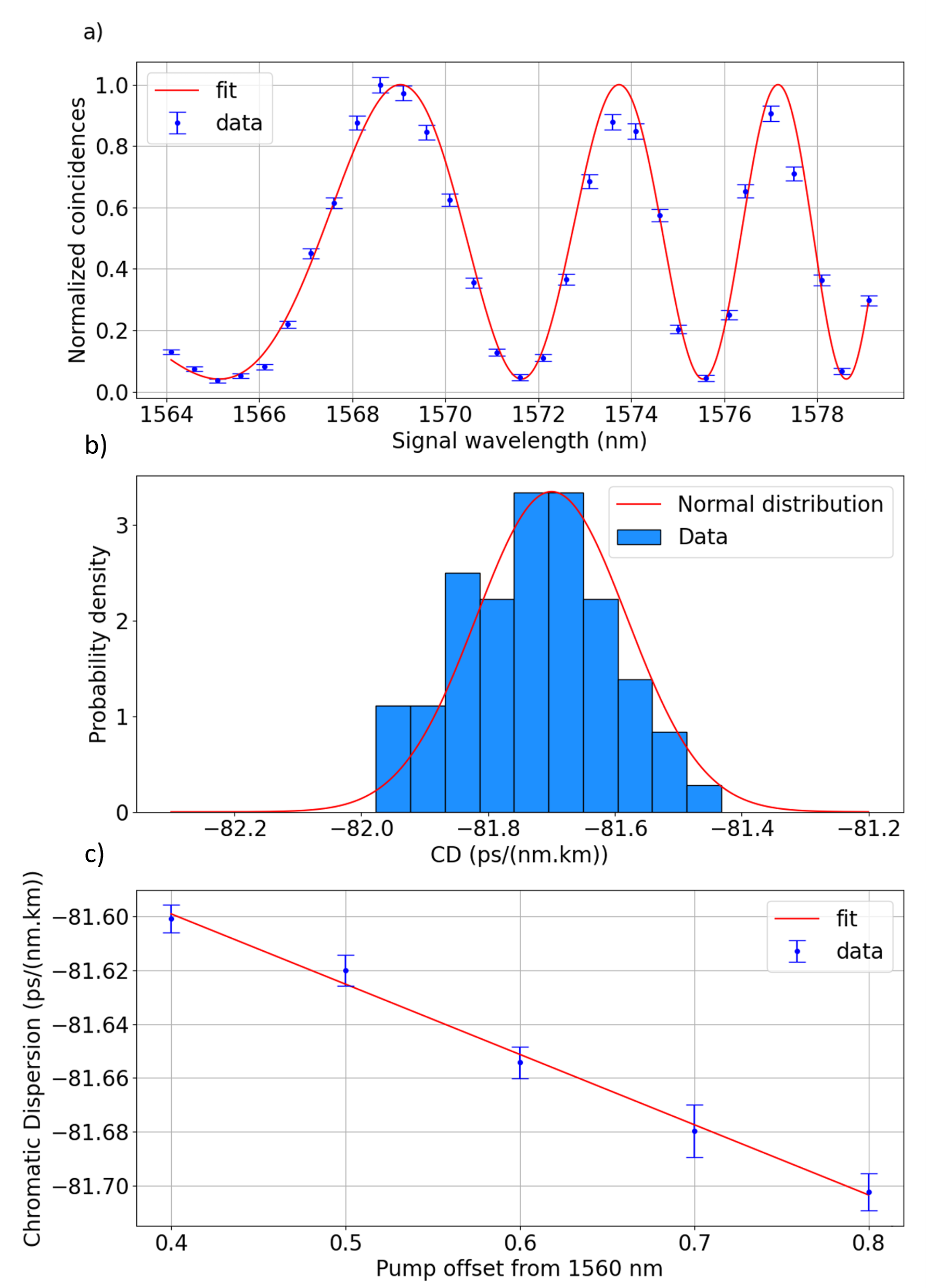}
    \caption{a) Normalized coincidence measurement as a function of the signal wavelength, the blue dots are fitted using Eq.~\ref{eq: coinc rate}. b) Histogram of extracted CD for 100 measurements at $\lambda_p=1560.800 \, \text{nm}$. The red line is calculated assuming a normal distribution. c) Extracted CD for different pump wavelengths. The red curve represents a linear fit used to extract the TOD.}
    \label{fig: CD and TOD}
\end{figure}

\textit{Measurement results.} The precision is estimated through the measurements of the normalized coincidence spectrum, as a function of the signal photon wavelength (corresponding to the first filter spectral position). An example is  presented in Fig.~\ref{fig: CD and TOD} (a) in blue. At each measurement point, filters with a spectral bandwidth of approximately $\Delta \lambda \approx 500 \, \text{pm}$ are adjusted in accordance with the energy conservation of the SPDC process around the pump wavelength. In our case, the maximum coincidence rate is approximately $5\, \text{kcts/s}$ to prevent statistical errors arising from the Poissonian emission statistic of photon pairs~\cite{PhysRevApplied.20.024026_dalidet}. The interferogram is normalized by the single photon rate, removing the cardinal sine term found in Eq.~\ref{eq: coinc rate}. Then, this interferogram is used to fit the experimental values (red curve) in order to extract the CD of the SUT. Since the statistics of $N$ distinct measurements follows a normal distribution, the experiment is replicated 100 times to ascertain the central value of the CD and its statistical precision, shown in Fig.~\ref{fig: CD and TOD} (b). For a pump wavelength of $\lambda_p = 1560.600\, \text{nm}$, we obtain a CD value of $-81.654(6)\, \text{ps/(km.nm)}$, with a statistical error of $7.10^{-3}\%$. To the best of our knowledge, this represents a state-of-the-art measurement for chromatic dispersion at a telecom wavelength, being one order of magnitude higher than classical and quantum measurements \cite{kaiser_quantum_2018}. We emphasize that the calibration of the CD of the system itself (without the SUT) is essential to exclusively extract the CD of the SUT. While in our case this measurement leads to an apparent negligible value (no visible fringes in the spectrum), measuring the CD subtraction of 2 SUTs with different lengths would lead to a calibration-free approach.\\
Then, we proceed with evaluating the accuracy by measuring the derivative of the CD, known as the third-order dispersion (TOD). The derivative allows us to infer the variation in CD, thus enabling to compare the slopes of the estimated and target parameters, the latter being provided by the manufacturer. The measurement procedure, involving the recording of a coincidence spectrum, was repeated for various pump wavelengths ranging from $\lambda_p = 1560.400 \, \text{nm}$ to $\lambda_p = 1560.800 \, \text{nm}$ which correspond to the phase matching bandwidth of the crystals, with a step of $\Delta\lambda_p = 100 \, \text{pm}$. The results are illustrated in Fig. \ref{fig: CD and TOD} (c). 

\textit{Result analysis.} Assuming that the second-order derivative of the CD, i.e., the fourth-order dispersion, is negligible, the TOD is easily extracted through a simple linear fit of the experimental CD values. We obtain a TOD value of $-0.26(1)\, \text{ps/(nm}^2\text{.km)}$, corresponding to a fit uncertainty of less than 5\%, which stands as a state-of-the-art measurement for TOD \cite{Grosz:14_first_order}. We emphasize that both the CD and TOD values fall within the range of the manufacturer's specifications~\cite{Thorlabs}. While the statistical error of the CD measurement is related to the precision of the protocol, the quadratic error of the extracted TOD indicates the accuracy of the measurement. Both values, directly accessible from the experimental parameters of the setup, attest to the quality of the results related to the chosen architecture and its implementation, ensuring the stability and reliability of the measurement procedure.

\noindent It is worth noting that measuring the spectrum of the system for different pump wavelength correspond to measuring its joint spectral intensity (JSI), which encodes the spectral correlations of the paired photons as a function of the pump wavelength. Thus, by acquiring the JSI with a spectrally broaden pump laser, the CD and therefore TOD could be extracted by a single-shot measurement. By assuming the pump bandwidth to match the phase matching bandwidth of the crystals, the limitation of this measurement would be determined by the latter \textit{i.e.}, using shorter crystals or aperiodic poled crystals would provide access to more CD values from which higher order dispersion terms could be inferred. Moreover, as the Sagnac configuration relaxes constraints on the pump coherence since the optical paths are identical in both CWi and CCWi directions, a standard wide-spectrum light source such as a superluminescent diode could be employed. This is in contrast to both Mach-Zehnder and Michelson configurations, which necessitate the use of a pulsed coherent laser.

\begin{figure}[!ht]
    \centering
    \includegraphics[width=0.9\linewidth]{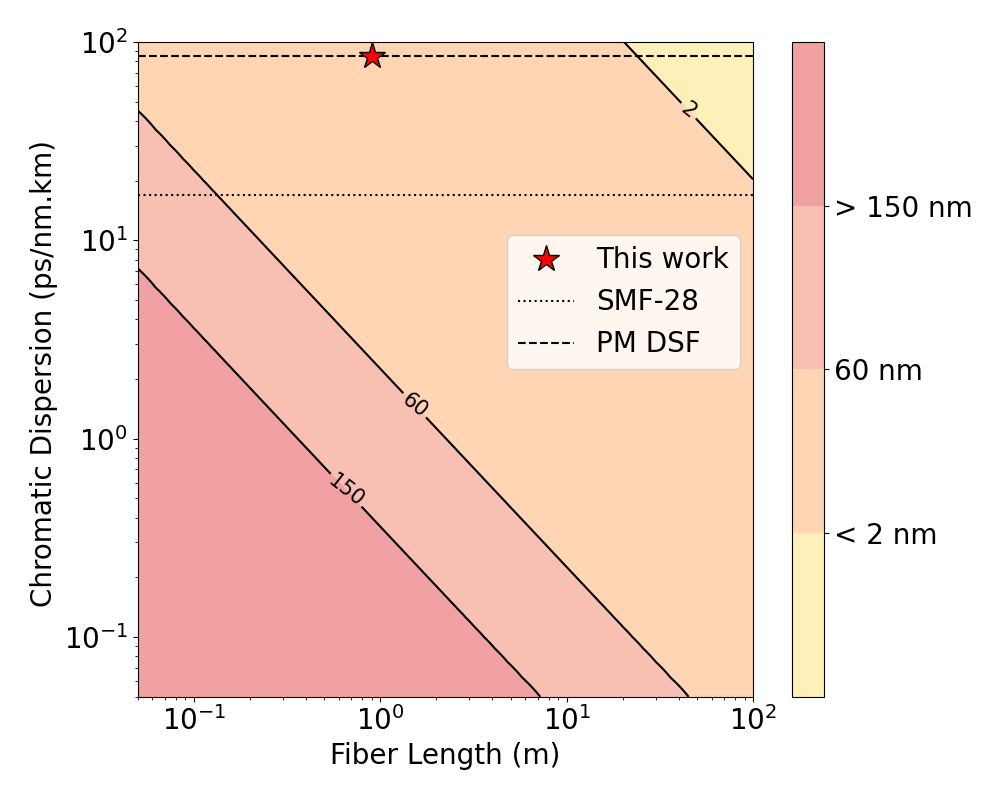}
    \caption{Spectral width of the first quadratic two-photon fringe as a function of the sample length and its CD value. The dotted lines represent the dispersion value of the SUT and a standard SMF-28 telecom fiber. PM DSF: polarization maintaining dispersion-shifted fiber.}
    \label{fig: CD limit}
\end{figure}

In a broader context, the working range of the non-linear Sagnac loop for CD measurement can be calculated by deducing the spectral width of the first quadratic fringe of the two-photon interference:

\begin{equation}\label{eq: fringe bandwidth}
    \Delta\lambda = \Bigl(\frac{\lambda_p^4}{2\pi c^2 |\beta^{(2)}|L}\Bigr)^{\frac{1}{2}} \, .
\end{equation}

In this equation, we assume $\phi_{off} = 0$. The spectral width is presented in Fig.~\ref{fig: CD limit} as a function of the SUT length and its CD. This synthetic figure can be divided into four zones:

\begin{itemize}
    \item [-] The yellow zone corresponds to scenarios where both dispersion and SUT length play the most significant role (resulting in narrow fringes), requiring extremely narrow filters. \vspace{-0.35 cm}
    \item [-] The orange zone represents the CD measurements accessible with our system, including the interferometer and filters.\vspace{-0.35 cm}
    \item [-] The salmon and red-colored zones depict cases where either the length of the SUT is short and/or it's CD value is low (resulting in wide fringes). To measure the CD in these scenarios, filters with a broader bandwidth are required, and the waveguides should be shorter to achieve a wider single-photon spectrum.
\end{itemize}
While our system is mainly limited by the bandwidth and operating range of the filters used for coincidence measurements, the orange zone remains fully accessible even for longer samples, given the self-stabilizing nature of the Sagnac loop. Consequently, CD can be extracted from relatively small samples, ranging from a few centimeters of optical materials to kilometer-long fibers, even when the latter approach zero-dispersion values, demonstrating the versatility of the method which, to the best of our knowledge, is not currently available.

\section{Conclusion}

In this work, we have introduced a novel quantum non-linear interferometric configuration based on a Sagnac loop, enabling access to an unattainable parameter: the optical phase in the frequency domain. This approach enables precise, accurate, and self-stabilized optical phase measurements, showcased by the measurement of CD. By exploiting the polarization entanglement of photon pairs within this non-linear Sagnac architecture, our method significantly surpasses conventional techniques in both accuracy and experimental simplicity. The system's intrinsic self-stabilizing nature, which eliminates the need for intricate active stabilization mechanisms, combined with the advantages of utilizing entangled photon pairs, constitutes one of its key strengths.\\
Beyond its robustness, the versatility of this interferometer — capable of characterizing a broad spectrum of optical samples, ranging from kilometers of fiber to materials only a few centimeters in length — underscores the maturity and practical potential of quantum technologies. Furthermore, this work paves the way for SUTure miniaturization efforts, particularly with the integration of $\chi^{(3)}$ non-linearities into compact, micrometer-scale devices, reinforcing the transformative potential of quantum metrology based on photon entanglement. Our results demonstrate a clear path forward for the development of photon-based sensors that marry high performance with user-friendly operation, ensuring that quantum-enhanced measurement systems will continue to push the frontiers of experimental science and technology.

\vspace{20pt}

\section*{Acknowledgment}

This work has been conducted within the framework of the project OPTIMAL granted by the European Union by means of the Fond Européen de développement régional (FEDER). The authors also acknowledge financial support from the Agence Nationale de la Recherche (ANR) through the projects METROPOLIS (ANR-19-CE47-0008), QAFEINE (21-ASTR-0007-DA), PARADIS (ANR-22-ASTR-0027-01), ADEQUADE (European Defense Fund, 2023).\\

\section*{Author information}

All the authors contributed equally to the entire process, from the first draft to the final version of the manuscript before submission. We all read, discussed, and contributed to the writing, reviewing, and editing. S.T. and L.L coordinated and managed the project, ensuring its successful completion.

\section*{Competing interests}
The authors declare that there are no competing interests.

\section*{Data Availability}
Data are available from the authors on reasonable request.

\end{document}